\documentclass[prl,twocolumn,preprintnumbers,tightenlines,superscriptaddress]{revtex4}
\usepackage{graphicx}
\usepackage{amssymb}
\usepackage{amsmath,mathrsfs,verbatim}
\usepackage{xcolor}
\usepackage{graphics}
\usepackage{times}
\usepackage{latexsym}
\newcounter{sec}

\begin{document}

\title{The Warm Dark Matter Doorframe for Light Dark Matter Direct Detection Experiments}

\author{Ran Huo}
\author{Tao Xu}
\affiliation{Zhejiang Institute of Modern Physics, Zhejiang University, Hangzhou, Zhejiang 310027, China}
\affiliation{Racah Institute of Physics, Hebrew University of Jerusalem, Jerusalem 91904, Israel}

\date{\today}

\begin{abstract}
If dark matter has even been in sufficient thermal contact with the visible sector and sufficiently light ($m_\chi\lesssim\mathcal{O}(10)~\text{keV}$), the thermal motion inherited from the visible sector will cause significant free streaming effect which is subject to the structure formation constraint, similar to the benchmark thermal warm dark matter model. Here we identify the interaction responsible for such thermal contact to be the interaction probed by the deep underground dark matter direct detection experiments. With the kinetic decoupling technique on the $m_\chi$ vs.~$\sigma$ plot we determine the bound shape in detail, and find that recasting the current Lyman-$\alpha$ bound gives a constraint of $m_\chi\gtrsim73~\text{keV}$, and it gets relaxed to $m_\chi\gtrsim35~\text{keV}$ for a smaller cross section of $\sigma<10^{-46}~\text{cm}^2$ with some model dependence. That can be taken as a generic ``no go'' constraint for light dark matter direct detection experiments, and the known caveats are if dark matter is axion-like with an early Bose-Einstein condensation form, or if there is Brownian motion protection of the free streaming.
\end{abstract}

\pacs{xxx}

\maketitle

\stepcounter{sec}
{\bf \Roman{sec}. Introduction.\;}
The nature of dark matter (DM) is believed to be beyond the standard model (SM) of particle physics. However, the $\Lambda$CDM model as the ``standard model'' in astrophysics and cosmology takes the existence of (cold) DM as a necessary part~\cite{Ade:2015xua}. Various experimental or observational efforts are made to probe the nature of DM beyond either the SM in particle physics or the $\Lambda$CDM in astrophysics. On the particle physics side, the deep underground DM direct detection experiment is undoubtedly a leading approach, which assumes DM has an interaction with baryon in a model-independent way, in the minimal form of elastic scattering cross section~\cite{Aprile:2018dbl,Cui:2017nnn,Akerib:2016vxi,Jiang:2018pic,Armengaud:2019kfj,Agnese:2017jvy}.

Another interesting kind on the astrophysics side is that if DM is not cold but warm, then the free streaming effect will erase the structure on small scales and get constrained~\cite{Irsic:2017ixq}. The warm DM (WDM) model represents a general class of models characterized by the thermal motion-induced free streaming of the bulk of DM, and if the DM is light enough then the free streaming tends to be significant. The thermal motion of DM is naturally obtained, given that a generic DM species has experienced the high-temperature stage in the early universe. If some portal interaction connecting to the visible sector characterized by the $T_\text{CMB}=2.725~\text{K}$ cosmic microwave background has ever established a kinetic equilibrium, a similar thermal motion of DM is guaranteed.

In this letter, we identify such interaction which gives thermal contact and free streaming effect to a light DM particle to be the interaction that gives the DM direction detection cross section. Then a generic light DM model with a direct detection cross section now becomes a WDM model and gets a constraint on its mass, which on the left side of the $m_\chi$ vs.~$\sigma$ plot excludes a mass region with an almost vertical edge. Parallel to the ``neutrino floor'' which is mostly horizontal~\cite{Cabrera:1984rr,Drukier:1986tm}, here we call such a constraint a ``warm dark matter doorframe'', implying that it gives a parameter region in which the DM direct detection should be accomplished inside.

The deep underground DM experiment is only probing the nonrelativistic limit of the DM nucleon interaction, while the thermal contact providing interaction should be fully relativistic. We will see that the promotion from the former to the latter introduces some model dependence. And in the main process to determine the thermal contact which is called the kinetic decoupling~\cite{Schmid:1998mx,Hofmann:2001bi,Profumo:2006bv,Bringmann:2006mu}, there are key differences from the previous studies, that the interested DM is light while the thermal source baryon is massive, and that both of the two have chemical potentials only with which the right number densities can be reproduced. At last, some mechanisms can escape or relax such constraints, and we can get similar constraints for other interacting particles such as electrons. In the following, we will go to them in detail.

\stepcounter{sec}
{\bf \Roman{sec}. The Warm Dark Matter Model.\;}
The benchmark thermal WDM model is a fermionic (using massive neutrino as a prototype) hot relic model, and in late time its number density determined by the hot relic nature multiplying the WDM rest mass determines its contribution to the critical density $\rho_c$, which is usually replacing the CDM contribution $\Omega_\chi$. When fixing $\Omega_\chi$, there is only one remaining parameter, which is usually taken as the WDM mass $m_\chi$. From it the WDM hot relic number density $n$ and the corresponding \emph{extrapolated} late time temperature $T_{\text{WDM}0}$ can be determined
\begin{equation}
n_0=\frac{\rho_c\Omega_\chi}{m_\text{WDM}},\qquad T_{\text{WDM}0}=\Big(\frac{\rho_c\Omega_\chi}{m_\text{WDM}}\frac{2\pi^2}{3\zeta(3)}\Big)^{\frac{1}{3}}.
\end{equation}
The late time virialization in structure formation gives the DM a galaxy dependent velocity which is much larger than this $T_{\text{WDM}0}$, but here by extrapolation we mean to naively use the $\propto a^{-1}$ relation to scale from an early time when such virialization is negligible and the particle is sufficiently relativistic. We use such an extrapolation way throughout the letter. Then the extrapolated DM characteristic free streaming velocity at today is
\begin{equation}\label{eq:FSv}
\langle v_0\rangle=\frac{7\pi^4}{180\zeta(3)}\Big(\frac{2\pi^2\rho_c\Omega_\chi}{3\zeta(3)m_\text{WDM}^4}\Big)^{\frac{1}{3}}
=
1.3\times10^{-8}\Big(\frac{5.3~\text{keV}}{m_\text{WDM}}\Big)^{\frac{4}{3}}c,
\end{equation}
where the factor $\frac{7\pi^4}{180\zeta(3)}\approx3.15$ is the ratio $\langle p\rangle/T$ for massless fermion. Early free streaming velocity is correcting this velocity by a factor $a^{-1}$ with $a$ the scale factor and normalization today $a_0=1$, and the widely used free streaming length~\cite{Kolb:1990vq} is the integration of the free streaming velocity in the local comoving frame from the bigbang till the matter-radiation equilibrium or even later (see Ref.~\cite{Huo:2019bjf} for a discussion of the validation for such integration).

\stepcounter{sec}
{\bf \Roman{sec}. From Direct Detection Cross Section to Matrix Element.\;}
Many DM models give very similar matrix elements for DM nucleon scattering. We illustrate with a fermionic DM $\chi$ and a vector boson mediator with mass $m_m$, and consider all combinations of vector current and axial current on both the dark side and the nucleon side (operator $\bar{\chi}(g_{V\chi}+g_{A\chi}\gamma^5)\gamma^\mu\chi\bar{n}(g_{Vn}+g_{An}\gamma^5)\gamma_\mu n$). Not concentrated on the nonrelativistic limit, the full matrix element with an average of initial state degree of freedom (DOF) and sum over final state DOF is
\begin{align}\label{eq:matrixelementSq}
&\frac{|\mathcal{M}|^2}{2(2S_\chi+1)}=\frac{16m_n^2m_\chi^2(g_{Vn}^2g_{V\chi}^2\hspace{-0.2em}+\hspace{-0.2em}3g_{An}^2g_{A\chi}^2)}{(q^2+m_m^2)^2}\\
\times&\bigg(1+\frac{p^2(m_n^2\hspace{-0.2em}+\hspace{-0.2em}m_\chi^2\hspace{-0.2em}+\hspace{-0.2em}2E_nE_\chi)\hspace{-0.2em}+\hspace{-0.2em}2p^4}{m_n^2m_\chi^2}
\frac{(g_{Vn}^2\hspace{-0.5em}+\hspace{-0.3em}g_{An}^2)(g_{V\chi}^2\hspace{-0.5em}+\hspace{-0.3em}g_{A\chi}^2)}{g_{Vn}^2g_{V\chi}^2+3g_{An}^2g_{A\chi}^2}\bigg),\nonumber
\end{align}
where $m_n$ is the nucleon mass, $q$ is the momentum transfer, $p$ is the momentum magnitude in the center of mass frame, and $E_{n,\chi}=(m_{n,\chi}^2+p^2)^{\frac{1}{2}}$. Then if DM is instead a complex scalar and on the dark side there can only be vector current coupling, the matrix element is obtained by setting the above $g_{A\chi}=0$. Also, if the mediator is not a vector but a scalar, the matrix element is obtained by dropping the terms proportional to the power of $p$. With such variants this example should cover the most popular cases in the DM-nucleon scattering operator list~\cite{Fan:2010gt,Fitzpatrick:2012ix}.

In Eq.~\ref{eq:matrixelementSq} we keep all terms which may be needed later in kinetic decoupling, but for the nonrelativistic DM direct detection, we only need the $p\to0$ limit. In such limit the matrix element square is related to the experimental spin-independent/dependent cross section via $\sigma_n=\frac{1}{16\pi(m_\chi+m_n)^2}\frac{|\mathcal{M}|^2}{2(2S_\chi+1)}\big|_{p=0}$ by
\begin{equation}\label{eq:SIandSD}
\sigma_n^\text{SI}=\frac{\mu_{n\chi}^2}{\pi}\frac{g_{Vn}^2g_{V\chi}^2}{(q^2+m_m^2)^2},~~
\sigma_n^\text{SD}=\frac{\mu_{n\chi}^2}{\pi}\frac{3g_{An}^2g_{A\chi}^2}{(q^2+m_m^2)^2},
\end{equation}
where $\mu_{n\chi}=\frac{m_nm_\chi}{m_n+m_\chi}$ is the DM-nucleon reduced mass, and the spin dependence is implicitly in the couplings $g_{An}$ and $g_{A\chi}$. For a certain $m_\chi$ and assuming one coupling (either $g_V$ or $g_A$) dominates on either the DM or the nucleon side, the unknown part of the $(g_{Vn}^2g_{V\chi}^2+3g_{An}^2g_{A\chi}^2)/(q^2+m_m^2)^2$ factor in the full matrix element square is proportional to $\sigma_n$ and can be replaced. Moreover, the terms in nonzero powers of $p$ can also be determined because the coupling mixing factor can be reduced under the previous assumption. In all, the full matrix element square is proportional to $\sigma_n$ and thus determined.

\stepcounter{sec}
{\bf \Roman{sec}. Kinetic Decoupling of Light Dark Matter From Baryon.\;}
In the expanding universe the momentum of thermal bath particle redshifts as $a^{-1}$, then other species which is kept in thermal equilibrium and shares the same characteristic momentum till decoupling has a degeneracy in determining the late time characteristic momentum: early or late decoupling gives the same late time free streaming momentum. The way to break such degeneracy relies on the entropy transfer in the thermal bath~\cite{Husdal:2016haj}. As the universe cools, many particle species are removed from the thermal bath content, due to either annihilation of the dominant symmetric part of particles and their antiparticles or the associated phase transition such as the QCD one. And the conserved entropy of the annihilated part is transferred into the remaining part, heating the remaining thermal bath and possibly the DM if it still has sufficient thermal contact. Such process can be summarized as $g_{\ast s}(T)T^3a^3=$constant, where $g_{\ast s}(T)$ is the effective DOF for entropy, and in the combination of SM and $\Lambda$CDM it monotonically decreases from an early value of $106.75=g_{\ast s,\text{SM}}$ when all the SM DOFs are in thermal bath to about $3.93=g_{\ast s0}$ as the late time/current value with contribution only from photons and neutrinos. Consequently for light species, ignoring other contributions to $g_{\ast s}(T)$ such as that from the dark sector and extrapolating to today again gives
\begin{equation}\label{eq:momentum}
T_{\chi 0}=T_\text{CMB}\Big(\frac{g_{\ast s0}}{g_{\ast s\text{dec}}}\Big)^{\frac{1}{3}}=2.3\times10^{-7}\Big(\frac{3.93}{g_{\ast s\text{dec}}}\Big)^{\frac{1}{3}}~\text{keV}.
\end{equation}

However, from the extrapolated temperature to the interested free streaming effect, there is still one subtlety, that the ratio $\langle p\rangle/T$ will not always be the same as the WDM one. This ratio is model dependent. For example, if DM is a light boson it will be $\frac{\pi^4}{30\zeta(3)}\approx2.70$. But for our interested case $T_\text{dec}\lesssim m_\chi$, $p^2\sim m_\chi T$, and the decoupling $\langle p\rangle/T$ ratio can be noticeably larger than the massless case value. Keeping this fact explicit and dividing $m_\chi$, we can compare to the Eq.~\ref{eq:FSv}, and to achieve the same marginal $\langle v_0\rangle$ we get the doorframe marginal value
\begin{equation}\label{eq:doorframe}
m_{\chi\text{DF}}=58\Big(\frac{\langle p\rangle}{T}\Big/\frac{7\pi^4}{180\zeta(3)}\Big)\Big(\frac{3.93}{g_{\ast s\text{dec}}}\Big)^{\frac{1}{3}}\Big(\frac{m_\text{WDM}}{5.3~\text{keV}}\Big)^{\frac{4}{3}}~\text{keV},
\end{equation}
Here we will use the simplified $\langle v_0\rangle$ criteria for the free streaming effect, and the exact matter power spectrum shape will be up to detailed calculation, with complete information of partition function. Except for the observational value of Lyman-$\alpha$ bound $m_\text{WDM}$, on the theory side we only need to determine the decoupling $g_{\ast s\text{dec}}$ and $\langle p\rangle/T$ for certain DM model parameter.

\begin{figure}[t!]
\includegraphics[scale=0.45]{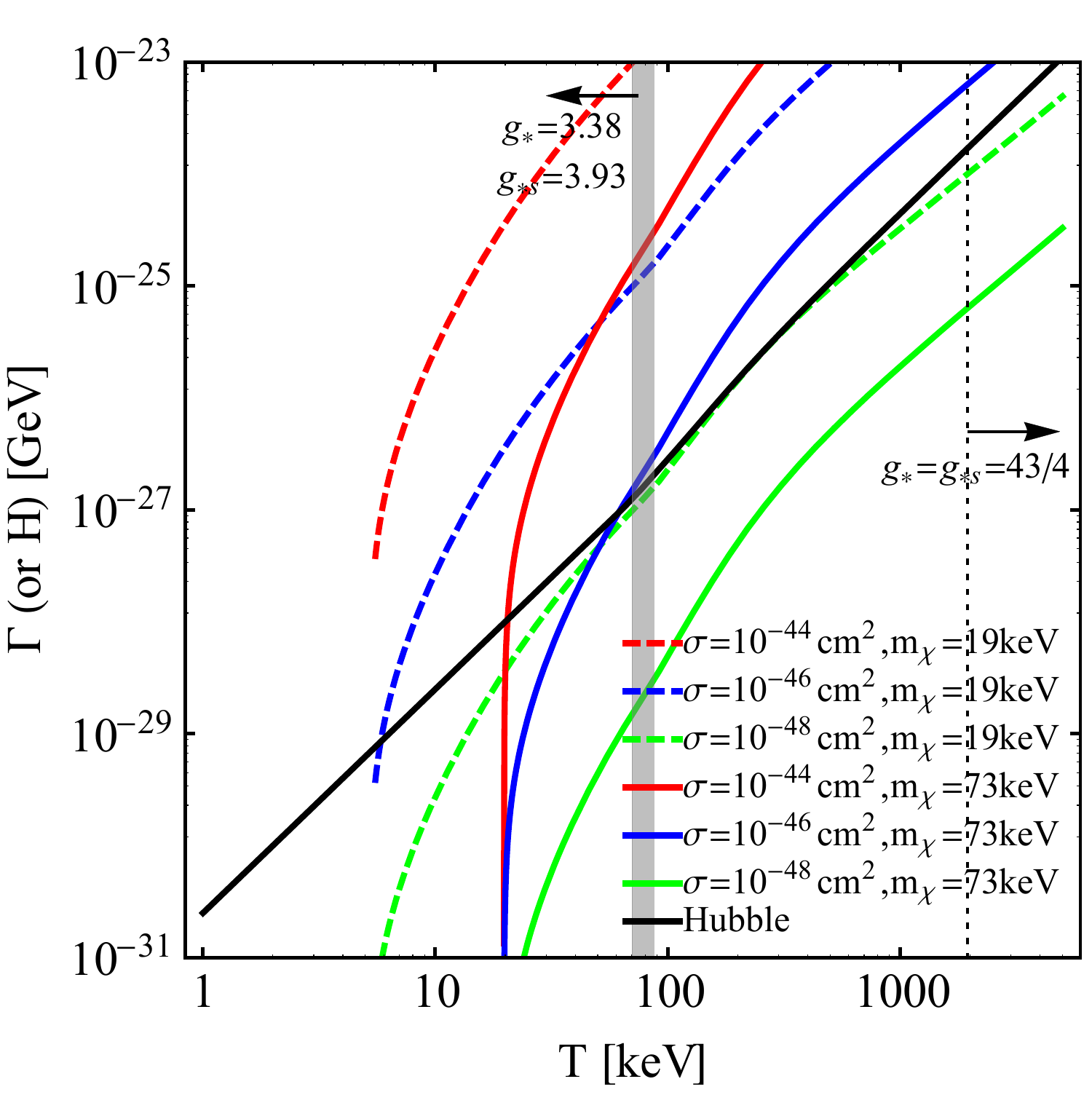}
\caption{Comparison of kinetic energy transfer rate $\Gamma$ (colored curves) to the Hubble parameter $H$, with the temperature in the multi-keV region. The inducing DM direct detection cross section is a spin-independent cross section with a vector mediator. The shown DM masses $19~\text{keV}$ (dashed curves) and $73~\text{keV}$ (colored solid curves) produce the same free streaming effect as the Lyman-$\alpha$ bounded $5.3~\text{keV}$ thermal WDM model when the kinetic decoupling happens at the effective DOF for entropy being $106.75$ and $3.93$, respectively (see Eq.~\ref{eq:upperlower}). For $\sigma\geq10^{-46}~\text{cm}^2$ the kinetic decoupling point as the $\Gamma$ and $H$ intersection has $g_{\ast s}=3.93$, meaning that then the $73~\text{keV}$ bound is self-consistent, rather than the $19~\text{keV}$.}
\label{fig:GammaVSHSI}
\end{figure}

Note that even before going into further detail, we already have a small $m_{\chi\text{DF}}$ range
\begin{equation}\label{eq:upperlower}
19\Big(\frac{m_\text{WDM}}{5.3~\text{keV}}\Big)^{\frac{4}{3}}~\text{keV}\leq m_{\chi\text{DF}}\leq 73\Big(\frac{m_\text{WDM}}{5.3~\text{keV}}\Big)^{\frac{4}{3}}~\text{keV}.
\end{equation}
Here we use the SM$+\Lambda$CDM range of $g_{\ast s}$. The left side is for early decoupling $g_{\ast s\text{dec}}=106.75$ and $\langle p\rangle/T=7\pi^4/(180\zeta(3))$ (ignore differences between a Fermi-Dirac distribution and a Bose-Einstein distribution), and the right side is for late decoupling $g_{\ast s\text{dec}}=3.93$ and $\langle p\rangle/T\approx4$. The last one is iteratively determined for a Maxwell-Boltzmann distribution and decoupling $m_\chi/T\approx3.6$ numerically when the Boltzmann suppression quickly becomes dominant (see Fig.~\ref{fig:GammaVSHSI}).

For the precise determination of the decoupling $g_{\ast s\text{dec}}$ and the corresponding $\langle p\rangle/T$, we need to study the kinetic decoupling process for the considered light DM. In contrast to the literature cases that DM is the nonrelativistic matter species and SM light particles (such as the electron) or even dark radiation are the effectively massless~\cite{Schmid:1998mx,Hofmann:2001bi,Profumo:2006bv,Bringmann:2006mu}, here the mass hierarchy is the opposite, that the DM is light and the baryon is relatively heavy. However, here we assume arbitrary mass for both DM and baryon and always use $E(q)=\sqrt{q^2+m^2}$. Moreover, the other key difference is that baryon as the thermal contact species has a number density which is not given by the thermal equilibrium distribution but determined observationally due to the cosmic baryon asymmetry. It is equivalent to chemical potential.

The detail of this calculation of the Boltzmann equation is given in the appendix. On the right hand side the basic idea is to expand around the trivial kinematics point $q'\to q$ of a $p+q\to p'+q'$ (as $n+\chi\to n'+\chi'$) configuration where the dominant contribution comes from, in both the Dirac delta function and the matrix element square Taylor expansion. Here we will directly give the final result with all the baryonic integration finished,
\begin{align}\label{eq:Boltzmann}
\frac{da^3f_\chi(q)}{a^3dt}&
=\sum_i\Big(\frac{\Gamma(\frac{i}{2}\hspace{-0.3em}+\hspace{-0.2em}3)(2m_nT_n)^{\frac{i+1}{2}}K_{\frac{i+5}{2}}({\textstyle\frac{m_n}{T_n}})}
{24\pi^{\frac{3}{2}}K_2({\textstyle\frac{m_n}{T_n}})}\frac{|\mathcal{M}|_i^2}{2S_\chi\hspace{-0.3em}+\hspace{-0.2em}1}\Big)\nonumber\\
&\hspace{-3em}\times\frac{n_{n0}g_{\ast s}(T_n)T_n^3}{E_\chi(q)^3g_{\ast s0}T_\text{CMB}^3}(E_\chi(q)T_n\vec{\partial}_q\vec{\partial}_q\hspace{-0.3em}+\hspace{-0.2em}\vec{q}~\vec{\partial}_q\hspace{-0.3em}+\hspace{-0.2em}3)f_\chi(q).
\end{align}
Here $n_{n0}\approx2.5\times10^{-7}~\text{cm}^{-3}$ is the baryon number density today, $|\mathcal{M}|_i^2$ is the coefficient of the Taylor series $|\mathcal{M}|^2=\sum_ip^i|\mathcal{M}|_i^2$, and $K_n$ is the modified Bessel function of the second kind.

The above Boltzmann equation can be multiplied (here $\frac{q^2}{3E(q)}$) and integrated to get the evolution of interested quantity (here the temperature), and the remaining factor is the rate $\Gamma$ of kinetic energy transfer and can be compared to the Hubble parameter $H$ for decoupling~\cite{Visinelli:2015eka}, which reads
\begin{align}\label{eq:Gamma}
\Gamma=&\sum_i\Big(\frac{\Gamma(\frac{i}{2}\hspace{-0.3em}+\hspace{-0.2em}3)(2m_nT_n)^{\frac{i+1}{2}}K_{\frac{i+5}{2}}({\textstyle\frac{m_n}{T_n}})}
{72\pi^{\frac{3}{2}}K_2({\textstyle\frac{m_n}{T_n}})}\frac{|\mathcal{M}|_i^2}{2S_\chi\hspace{-0.3em}+\hspace{-0.2em}1}\Big)\frac{g_{\ast s}(T_n)T_n^3}{g_{\ast s0}T_\text{CMB}^3}\nonumber\\
\times&\frac{n_{n0}}{m_\chi^2T_\chi^2K_2({\textstyle\frac{m_\chi}{T_\chi}})}\int\frac{dqq^4e^{-\frac{E_\chi(q)}{T_\chi}}}{E_\chi(q)^4}\Big(3-\frac{q^2}{E_\chi(q)T_\chi}\Big).
\end{align}

\begin{figure}[t!]
\includegraphics[scale=0.45]{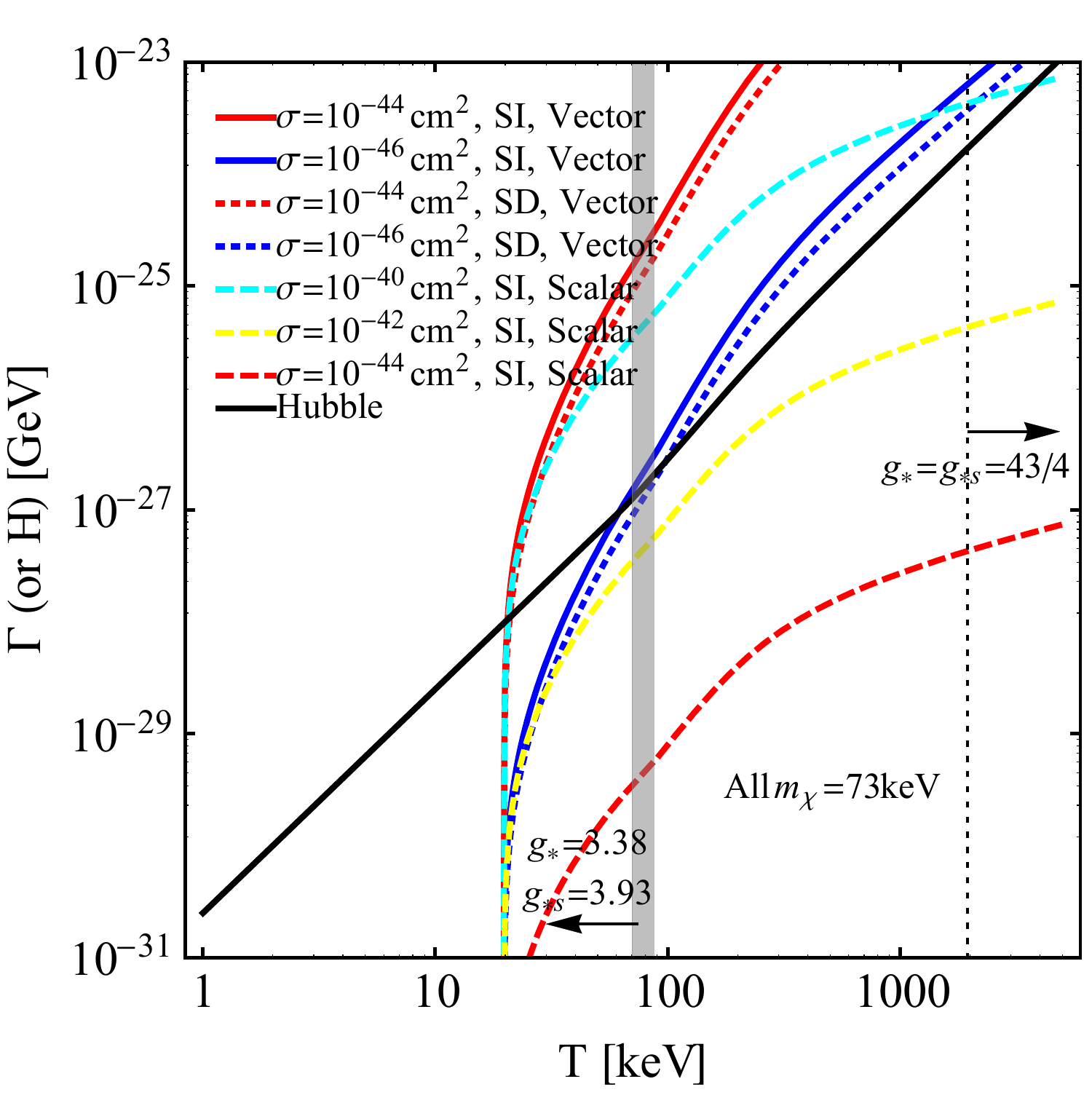}
\caption{Same as Fig.~\ref{fig:GammaVSHSI}, but for comparison with other DM direct detection model choices such as spin dependent cross section or scalar mediator, for a fixed $m_\chi=73~\text{keV}$.}
\label{fig:GammaVSHAll}
\end{figure}

\stepcounter{sec}
{\bf \Roman{sec}. Results.\;}
In Fig.~\ref{fig:GammaVSHSI} and \ref{fig:GammaVSHAll} we plot such comparison regardless of DM being fermion or scalar, with the former focusing on the spin independent cross section with a massive vector mediator and showing two illustrative masses, and the latter focusing on the comparison between the spin dependence as well as the difference between vector mediator and scalar mediator, but for a fixed mass. On each plot, the kinetic decoupling corresponds to the intersection point of $\Gamma$ and $H$ curves. Taking the lowest $g_{\ast s}=3.93$ for example, according to Eq.~\ref{eq:doorframe} it corresponds to a mass of $m_\chi=73~\text{keV}$, and if with a given cross section $\sigma$ it has a kinetic decoupling temperature $T$ which gives $g_{\ast s}(T)=3.93$, then such $\sigma$ is the self-consistent bound for this $m_\chi$. From Fig.~\ref{fig:GammaVSHSI} we can see a sharp Boltzmann cutoff of the rate $\Gamma$. We find that approximately the cutoff temperature is always at $T\approx m_\chi/3.6$ for the interested DM mass, which usually dictates the kinetic decoupling. In the interested series of solid curves of $m_\chi=73~\text{keV}$, a cross section of $\sigma\geq10^{-46}~\text{cm}^2$ will give such intersection, so the WDM doorframe bound will be $m_\chi>73~\text{keV}$ for $\sigma\geq10^{-46}~\text{cm}^2$, which could potentially be covered by future light dark matter direct detection experiments~\cite{Schutz:2016tid,Knapen:2016cue,Knapen:2017ekk,Griffin:2018bjn,Hertel:2018aal,Essig:2019kfe}.

\begin{figure}[t!]
\includegraphics[scale=0.45]{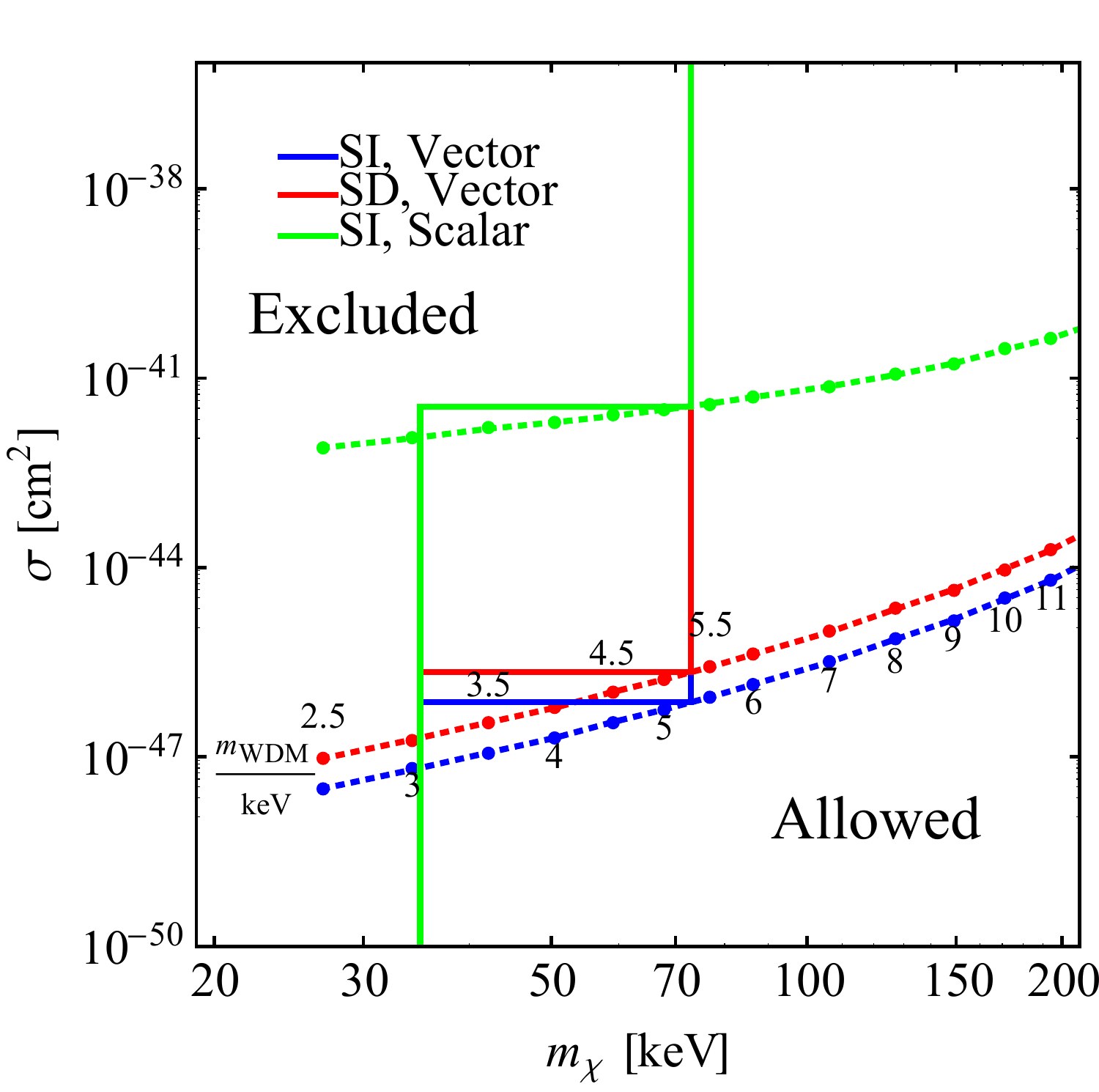}
\caption{The WDM doorframe bounds for light DM direct detection on the $m_\chi$ vs.~$\sigma$ plane, with thick solid lines recasting the current $2\sigma$ confidence level Lyman-$\alpha$ bound of $m_\chi=5.3~\text{keV}$, for three model choices. On the other hand, each dashed curve gives the trajectory of the corner position of the doorframe bound, for any perspective Lyman-$\alpha$ WDM bound giving the horizontal transition $\sigma$. For any observational $m_\chi$ value, the upper part of doorframe bound is $m_\chi>73(\frac{m_\text{WDM}}{5.3~\text{keV}})^{\frac{4}{3}}~\text{keV}$ and the lower part is $m_\chi>35(\frac{m_\text{WDM}}{5.3~\text{keV}})^{\frac{4}{3}}~\text{keV}$, respectively.}
\label{fig:DoorFinal}
\end{figure}

Then for a little bit smaller $\sigma$, the $\Gamma=H$ kinetic decoupling intersection will moves to higher temperature which corresponds to larger $g_{\ast s}$, and smaller $m_\chi$ for a fixed $v_0$ according to Eq.~\ref{eq:doorframe}, closer to the dashed curves of $m_\chi=19~\text{keV}$ on Fig.~\ref{fig:GammaVSHSI}. However, we can see that the $\Gamma$ curves quickly become parallel to the $H$ one, and there will be no intersection practically. Here we will always assume that the behavior of $\Gamma$ becomes parallel to that of $H$ for a higher $T$ right after $\sigma$ drops below the cross section minimum for an intersection, and such approximation will result in a horizontal leftward weakening of the doorframe bound on the $m_\chi$ vs.~$\sigma$ plot, which should be a working approximation. On the other hand, $\Gamma$ is proportional to the true baryon number density, so it is suppressed by $\eta=n_b/n_\gamma\approx6\times10^{-10}$~\cite{Ade:2015xua} or so compared to a species before the annihilation of the symmetric part. When the temperature is slightly below $150$~MeV~\cite{Husdal:2016haj}, in the thermal bath there is still the symmetric SM pion DOF contribution, and this is the last point during the cosmic cooling when the unsuppressed symmetric contribution still thermally connecting DM and nucleon (with implicit assumption such as the DM and nucleon interaction has a quark/gluon level origin, which is similar for a SM pion). Naively boosting $\Gamma$ by these ten orders from $\eta^{-1}$ then in this temperature region the $\Gamma=H$ decoupling should be easily reached, for a $\sigma$ below the above last section minimum. Here we will use $g_{\ast s\text{dec}}=69/4$ corresponding to the presence of SM $\pi^\pm$ and $\pi^0$, and correspondingly $m_{\chi\text{DF}}=35~\text{keV}$ where we have used $\frac{\langle p\rangle}{T}=\frac{7\pi^4}{180\zeta(3)}$ since then $T\gg m_\chi$. Here we are not interested in the lower side extension of the cross section due to the neutrino floor. It completes the final doorframe bound contour for a spin-independent cross section with a massive vector mediator, which is shown in Fig.~\ref{fig:DoorFinal}.

Other particle physics models will lead to slightly different contours, due to differences in the horizontal transition cross section. In the $p^i$ series expansion of the matrix element square, the $p^2$ term contributes dominantly to Eq.~\ref{eq:Gamma} for light DM. We can estimate the ratio of the $p^2$ term to the $p^0$ term contribution to $\Gamma$ to be $\frac{2m_nT(m_n^2+m_\chi^2)}{m_n^2m_\chi^2}\approx\frac{2m_nT}{m_\chi^2}$ which can be order $10^4$. In the scalar mediator case without such term, the cross section needs to be larger by the corresponding 4 orders to give the same $\Gamma$; and even in the vector mediator case for a spin-dependent cross section the factor $3$ in $3g_{An}^2g_{A\chi}^2$ leads to a relatively suppression in the $p^2$ term, and the required cross section will be slightly larger. The induced difference in $\Gamma$ can be seen in Fig.~\ref{fig:GammaVSHAll}.

Finally in Fig.~\ref{fig:DoorFinal} the extracted doorframe bounds are plotted for all model choices. In addition to the current bound when recasting $m_\text{WDM}=5.3~\text{keV}$~\cite{Irsic:2017ixq}, we draw the trajectory of the corner for a range of $m_\text{WDM}$, which serves more conservative estimation or future improved observation.

\stepcounter{sec}
{\bf \Roman{sec}. Caveat and Other Discussion.\;}
Except for the trivial possibility that the DM is heavy, so the free streaming is diminished, other mechanisms that could avoid such a free streaming effect can be that the collisional thermalized DM is not the bulk but only a very subdominant fraction. The clear benchmark is the DM dominated by the vacuum misalignment produced axion. In the relevant early universe for $T>\mathcal{O}(100)~\text{MeV}$ what becomes the bulk of the present axion DM is still in the form of vacuum expectation value and cannot be scattered~\cite{Marsh:2015xka}. Alternatively, it can be that the self scattering keeps the DM in Brownian motion~\cite{Huo:2019bjf}. The widespread scattering disables any macroscopic free streaming.

On the other hand, some of the other frequent set up to cool the dark sector and escape constraint seems not working easily. For example, DM cannot get cooled in the dark sector due to a replication process, since the kinetic decoupling temperature is even smaller than its mass. Moreover, such constraints should not be interpreted as for thermally produced DM only. For example, if DM is the decay product of another particle then it will also obey this bound (even for the more massive mother particle).

Similarly, we could get a doorframe bound on the DM-electron scattering for future electron recoil experiments~\cite{Hochberg:2015pha,Hochberg:2016ntt,Derenzo:2016fse,Hochberg:2017wce,Kurinsky:2019pgb,Hochberg:2019cyy}. With number density $n_e=n_n$ the part for large cross section should be exactly the same, and the difference are the $m_{\chi\text{DF}}$ position of the small cross section part as well as the corner trajectory, for example using $g_{\ast s\text{dec}}=43/4$ for the presence of SM $e^\pm$ we get $m_{\chi\text{DF}}=41(\frac{m_\text{WDM}}{5.3~\text{keV}})^{\frac{4}{3}}~\text{keV}$.

\onecolumngrid
\appendix

\section{Appendix}

Here we give the details of the Boltzmann equation calculation, which is similar to~\cite{Bringmann:2006mu} but with some difference and clarification. In this letter we will always use Maxwell-Boltzmann distribution instead of a Bose-Einstein one or Fermi-Dirac one, which has simple integration expression with arbitrary mass. Therefore the partition function is $f_i(p)=e^{-\frac{E_i(p)-\mu_i}{T_i}}$ where $i=\chi$ or $n$, $E_i(p)=\sqrt{m_i^2+p^2}$ and $\mu_i$ is the chemical potential to reproduce the true number density for each species. The Boltzmann equation for a $p_\mu+q_\mu\to p'_\mu+q'_\mu$ as $n+\chi\to n'+\chi'$ configuration is
\begin{equation}
\frac{1}{a^3}\frac{d}{dt}(a^3f_\chi(q))=\frac{1}{2E_\chi(q)}\int\frac{d\vec{p}d\vec{p}'d\vec{q}'(2\pi)^4\delta^{(4)}(p'+q'-p-q)}{(2\pi)^32E_n(p)(2\pi)^32E_n(p')(2\pi)^32E_\chi(q')}
\frac{|\mathcal{M}|^2}{2S_\chi+1}J,
\end{equation}
where $J=(-f_n(p)f_\chi(q)+f_n(p')f_\chi(q'))$ is the notation in Ref.~\cite{Bringmann:2006mu}. Now the Taylor expansions of the momentum components and the energy component of Dirac delta function around $q'\to q$ are respectively
\begin{align*}
\delta(\vec{p}'+\vec{q}'-\vec{p}-\vec{q})=\delta((\vec{q}'-\vec{q})+(\vec{p}'-\vec{p}))&=\sum_{i=0}^\infty\frac{1}{i!}\big[(\vec{p}'-\vec{p})\vec{\partial}_{q'}\big]^i\delta(\vec{q}'-\vec{q}),\\
\delta(E_n(p')+E_\chi(q')-E_n(p)-E_\chi(q))
&=\sum_j\frac{1}{j!}\bigg[\frac{(E_\chi(q')-E_\chi(q))\partial}{\partial E_n(p')}\bigg]^j\delta(E_n(p')-E_n(p))\\
&=\sum_j\frac{1}{j!}\bigg[(E(q')-E(q))\frac{E(p')}{p'}\frac{\partial}{\partial p'}\bigg]^j\frac{E(p')}{p'}\delta(p'-p).
\end{align*}
Here we have used $E(p)dE(p)=pdp$, and $\delta(f(x)-f(x_0))=\frac{1}{|f'(x_0)|}\delta(x-x_0)$ to change the energy component into a Dirac delta function of $p'=|\vec{p}'|$. The derivatives in the Taylor expansion are moved using integration by parts, then the momentum component Dirac delta function can be trivially integrated out, and the other two differentiation are expanded in spherical system, with the azimuth always integrated out trivially. Now the RHS is
\begin{equation}
\text{RHS}=\int\sum_{0=j\leq i}^\infty\frac{(-1)^{i+j}\delta(p'-p)}{16i!j!E_n(p)E_\chi(q)}\bigg(
\bigg[(\vec{p}'-\vec{p})\cdot\vec{\partial}_{q'}\bigg]^i
\bigg[(E(q')-E(q))\frac{E(p')}{p'}\partial_{p'}\bigg]^j\frac{p^2dpd\cos\theta_pp'dp'd\cos\theta_{p'}}{(2\pi)^3E_\chi(q')}\frac{\sum|\mathcal{M}|^2}{2S_\chi+1}
J\bigg)\bigg|_{\vec{q}'\to\vec{q}}.
\end{equation}
Note that under the final $\vec{q}'\to\vec{q}$ condition imposed by the integrated out $\delta(\vec{q}'-\vec{q})$, both the last $J$ factor (together with the similar $p'\to p$ condition later) and the factor $(E(q')-E(q))$ will vanish eventually, unless each of them is acted by at least a derivative from the Taylor expansion. Namely for $J$ factor either a $\vec{\partial}_{q'}$ or a $\partial_{p'}$, and for each $(E(q')-E(q))$ a $\vec{\partial}_{q'}$. It turns out that in the Taylor expansion of both momentum delta and the energy delta we only need to keep up to the quadratic terms.

On the other hand, the $(\vec{p}'-\vec{p})$ is not a vanishing factor imposed by the delta function. In this 2 to 2 scattering configuration the fixed incident DM momentum $\vec{q}$ determines the $\theta=0$ direction, so $\vec{p}'-\vec{p}=p'\cos\theta_{p'}-p\cos\theta_p$ and gives factor of $\cos\theta$. Including the differentiation $d\cos\theta_pd\cos\theta_{p'}$ every term need odd power of both $\cos\theta_p$ and $\cos\theta_{p'}$ for it to be nonvanishing, so for a single $(\vec{p}'-\vec{p})$ we need another factor of $\cos\theta$, which can only come from the expansion of matrix element square. For a 2 to 2 scattering the matrix element square $|\mathcal{M}|^2$ is a function of the two (out of three) independent Mandelstam variables, which are chosen as $s$ and $t$ here. In this configuration $p_\mu=(\sqrt{p^2+m_n^2},~\vec{p})$ and $q_\mu=(\sqrt{q^2+m_\chi^2},~\vec{q})$, so $s=m_n^2+m_\chi^2+2E_n(p)E_\chi(q)-2pq\cos\theta_p$, and $t$ does not have a direct $\cos\theta_p$ dependence and is irrelevant. That means the $\cos\theta_p$ dependence is always in the Taylor expansion form of
\begin{align}
|\mathcal{M}|^2=&|\mathcal{M}|^2\Big|_{\cos\theta_p=0}+\frac{\partial}{\partial(2E_n(p)E_\chi(q))}|\mathcal{M}|^2\Big|_{\cos\theta_p=0}(-2pq\cos\theta_p)+\cdots\\
=&|\mathcal{M}|^2\Big|_{\cos\theta_p=0}+\frac{E_n(p)}{p}\frac{\partial}{\partial p}|\mathcal{M}|^2\Big|_{\cos\theta_p=0}\Big(-\frac{pq}{E_\chi(q)}\cos\theta_p\Big)+\cdots.
\end{align}
The terms linear in $(\vec{p}'-\vec{p})$ will be combined with the second term in the above expansion, and the terms quadratic will be combined with the first. The $\partial_p$ acting on $|\mathcal{M}|^2$ is similarly integrated by part in the following. Note that this expansion in $\cos\theta_p$ here is independent of the expansion of power of $p$ in the main text.

Now we can collect all the nonvanishing terms. Some of the repeated elementary calculation are
\begin{itemize}
\item $\partial_{p'}J|_{\vec{q}'\to\vec{q},p'\to p}=-\frac{p'}{T_nE_n(p')}f_n(p')f_\chi(q')|_{\vec{q}'\to\vec{q},p'\to p}=-\frac{p}{T_nE_n(p)}f_n(p)f_\chi(q)$ 
\item $\vec{\partial}_{q'}J|_{\vec{q}'\to\vec{q},p'\to p}=-\frac{\vec{q}'}{T_\chi E_\chi(q')}f_n(p')f_\chi(q')|_{\vec{q}'\to\vec{q},p'\to p}=-\frac{\vec{q}}{T_\chi E_\chi(q)}f_n(p)f_\chi(q)$
\item $\vec{\partial}_{q'}E_\chi(q')|_{\vec{q}'\to\vec{q}}=\frac{\vec{q}}{E_\chi(q)}$, $\vec{\partial}_{q'}\vec{\partial}_{q'}E_\chi(q')|_{\vec{q}'\to\vec{q}}\approx\frac{3}{E_\chi(q)}$.
\item $\int d\cos\theta_pd\cos\theta_{p'}(p'\cos\theta_{p'}-p\cos\theta_p)\frac{-E_n(p)q}{E_\chi(q)}\cos\theta_p=pq\frac{E_n(p)}{E_\chi(q)}\int d\cos\theta_pd\cos\theta_{p'}\cos\theta_p^2=\frac{4}{3}pq\frac{E_n(p)}{E_\chi(q)}$
\item $\int d\cos\theta_pd\cos\theta_{p'}(p'\cos\theta_{p'}-p\cos\theta_p)^2=\frac{4}{3}(p^2+p'^2)$
\end{itemize}
Then after doing the $\delta(p'-p)dp'$ trivially, eventually to the same order as Ref.~\cite{Bringmann:2006mu}, we get all the nonvanishing terms
\begin{align*}
&\int\frac{dp}{12(2\pi)^3}\hspace{-1em}&&(-1)\hspace{-1em}&&\frac{\partial_p|\mathcal{M}|^2}{(2S_\chi+1)E_n(p)E_\chi(q)^2}p^3p\vec{q}\frac{E_n(p)}{E_\chi(q)}\vec{\partial}_q&&\hspace{-1em}f_n(p)f_\chi(q)\qquad\qquad&&i=1J&&j=0&\\
+&\int\frac{dp}{12(2\pi)^3}\hspace{-1em}&&(+1)\hspace{-1em}&&\frac{\partial_p|\mathcal{M}|^2}{(2S_\chi+1)E_n(p)E_\chi(q)^2}p^3p\vec{q}\frac{E_n(p)}{E_\chi(q)}\frac{\vec{q}E_n(p)}{pE_\chi(q)}\partial_p&&\hspace{-1em}f_n(p)f_\chi(q)&&i=1E&&j=1J&\\
+&\int\frac{dp}{12(2\pi)^3}\hspace{-1em}&&\Big(\frac{+1}{2}\Big)\hspace{-1em}&&\frac{|\mathcal{M}|^2}{(2S_\chi+1)E_n(p)E_\chi(q)^2}p^32p^2\vec{\partial}_q\vec{\partial}_q&&\hspace{-1em}\Big(\hspace{-0.3em}-\frac{T_nE_n(p)}{p}\partial_p\Big)f_n(p)f_\chi(q) &&i=2JJ&&j=0&\\
+&\int\frac{dp}{12(2\pi)^3}\hspace{-1em}&&\Big(\frac{-2}{2}\Big)\hspace{-1em}&&\frac{|\mathcal{M}|^2}{(2S_\chi+1)E_n(p)E_\chi(q)^2}p^32p^2\frac{\vec{q}E_n(p)}{pE_\chi(q)}\partial_p\vec{\partial}_q&&\hspace{-1em}f_n(p)f_\chi(q) &&i=2JE&&j=1J&\\
+&\int\frac{dp}{12(2\pi)^3}\hspace{-1em}&&\Big(\frac{-1}{2}\Big)\hspace{-1em}&&\frac{|\mathcal{M}|^2}{(2S_\chi+1)E_n(p)E_\chi(q)^2}p^32p\frac{3E_n(p)}{E_\chi(q)}\partial_p&&\hspace{-1em}f_n(p)f_\chi(q) &&i=2EE&&j=1J&\\
+&\int\frac{dp}{12(2\pi)^3}\hspace{-1em}&&\Big(\frac{-2}{2}\Big)\hspace{-1em}&&\partial_p\Big[\frac{|\mathcal{M}|^2}{(2S_\chi+1)E_n(p)E_\chi(q)^2}p^32p^2\frac{\vec{q}E_n(p)}{pE_\chi(q)}\Big]\vec{\partial}_q&&\hspace{-1em}f_n(p)f_\chi(q) &&i=2JE&&j=1r&\\
+&\int\frac{dp}{12(2\pi)^3}\hspace{-1em}&&\Big(\frac{+2}{2\cdot2}\Big)\hspace{-1em}&&\frac{|\mathcal{M}|^2}{(2S_\chi+1)E_n(p)E_\chi(q)^2}p^32p^2\Big(\frac{\vec{q}E_n(p)}{pE_\chi(q)}\Big)^2\partial_p\partial_p&&\hspace{-1em}f_n(p)f_\chi(q) &&i=2EE&&j=2JJ&\\
+&\int\frac{dp}{12(2\pi)^3}\hspace{-1em}&&\Big(\frac{+2\cdot2}{2\cdot2}\Big)\hspace{-1em}&&\partial_p\Big[\frac{|\mathcal{M}|^2}{(2S_\chi+1)E_n(p)E_\chi(q)^2}p^32p^2\Big(\frac{\vec{q}E_n(p)}{pE_\chi(q)}\Big)^2\Big]\partial_p&&\hspace{-1em}f_n(p)f_\chi(q) &&i=2EE&&j=2Jr&
\end{align*}
In the last two columns we record the orders of $i$ and $j$ in the momentum and energy Taylor expansion, and to what each of the induced derivatives are acted ($J$, or $(E(q')-E(q))$ labeled as $E$, or the rest labeled as $r$). The second row have the combination of the factor $\frac{(-1)^{i+j}}{i!j!}$ and the permutation factor of the action of the $\partial_{p'}$ and $\vec{\partial}_{q'}$ on each part. The 4th and 6th lines actually can be combined into a total derivative and cancel with each other, similar the 7th line partially cancels with the 8th line. In lines where there are $\vec{q}\vec{q}$ factor, one $\vec{q}$ is traded into $-\frac{\vec{q}}{T_\chi E_\chi(q)}$ upon acting on $f_n(p)f_\chi(q)$, while the $\partial_p$ in line 2, 5, and 7 and 8 (just one $\partial_p$) are similarly traded into $-\frac{p}{T_n E_n(p)}$. Eventually for $T_n=T_\chi$ which is true around the interested kinetic decoupling, the sum gives
\begin{align*}
\frac{1}{a^3}\frac{d}{dt}(a^3f_\chi(q))=&\int\frac{dp}{12(2\pi)^3E_\chi(q)^3}\partial_p\Big(\frac{p^4|\mathcal{M}|^2}{2S_\chi+1}\Big)(T_nE_\chi(q)\vec{\partial}_q\vec{\partial}_q+\vec{q}\vec{\partial}_q+3)f_n(p)f_\chi(q)\\
=&\int\frac{dp}{12(2\pi)^3E_\chi(q)^3}\sum_i\frac{(4+i)p^{3+i}|\mathcal{M}|^2_i}{2S_\chi+1}(T_nE_\chi(q)\vec{\partial}_q\vec{\partial}_q+\vec{q}\vec{\partial}_q+3)e^{-\frac{E_n(p)-\mu_n}{T_n}}
e^{-\frac{E_\chi(q)-\mu_\chi}{T_\chi}}
\end{align*}
We can see that even for arbitrary mass of both $n$ and $\chi$, the final expression takes a neat form.

In the integration over $dp$, we have made use of the integration identity of the modified Bessel function $\int_0^\infty e^{-\frac{\sqrt{m^2+p^2}}{T}}p^ndp=\frac{2}{n+1}\frac{\Gamma(\frac{n+3}{2})}{\sqrt{\pi}}(2mT)^{\frac{n}{2}}mK_{\frac{n}{2}+1}({\textstyle\frac{m}{T}})$. The remaining chemical potential for baryon can be solved with the observational baryon number density of the current universe, namely at an early temperature $T$,
$n_n(T)=\frac{n_{n0}}{a^3}=n_{n0}(\frac{g_{\ast s}(T)}{g_{\ast s0}})(\frac{T}{T_{\text{CMB,0}}})^3=2\int\frac{d^3p}{(2\pi)^3}e^{-\frac{\sqrt{m_n^2+p^2}-\mu_n}{T_n}}=\frac{m_n^2T_n}{\pi^2}K_2({\textstyle\frac{m_n}{T_n}})e^{\frac{\mu_n}{T_n}}$. Eventually we get nothing but Eq.~\ref{eq:Boltzmann}.

Upon the further integration of $d^3q$ with factor $\frac{q^2}{3E_\chi(q)}$, the left hand side still gives the temperature for arbitrary $\chi$ mass in Boltzmann distribution, which is seen with the integration identity of the modified Bessel function $\int_0^\infty e^{-\frac{\sqrt{m^2+p^2}}{T}}\frac{p^ndp}{\sqrt{m^2+p^2}}=\frac{\Gamma(\frac{n+1}{2})}{\sqrt{\pi}}(2mT)^{\frac{n}{2}}K_{\frac{n}{2}}({\textstyle\frac{m}{T}})$. Namely since
\begin{align}
&\frac{n_\chi}{2S_\chi+1}=\int\frac{d^3q}{(2\pi)^3}e^{-\frac{\sqrt{m_\chi^2+q^2}-\mu_\chi}{T_\chi}}=\frac{1}{2\pi^2}\frac{2}{3}\frac{\Gamma(\frac{5}{2})}{\sqrt{\pi}}(2m_\chi T_\chi)m_\chi K_2({\textstyle\frac{m_\chi}{T_\chi}})e^{\frac{\mu_\chi}{T_\chi}},\\
\intertext{we can see that}
&\int\frac{d^3q}{(2\pi)^3}e^{-\frac{\sqrt{m_\chi^2+q^2}-\mu_\chi}{T_\chi}}\frac{q^2}{3\sqrt{m_\chi^2+q^2}}=\frac{1}{2\pi^2}\frac{1}{3}\frac{\Gamma(\frac{5}{2})}{\sqrt{\pi}}(2m_\chi T_\chi)^2 K_2({\textstyle\frac{m_\chi}{T_\chi}})e^{\frac{\mu_\chi}{T_\chi}}=\frac{n_\chi}{2S_\chi+1}T_\chi.
\end{align}
For the right hand side, to leading order $(T_nE_\chi(q)\vec{\partial}_q\vec{\partial}_q+\vec{q}\vec{\partial}_q+3)e^{-\frac{E_\chi(q)-\mu_\chi}{T_\chi}}=\big(3-\frac{q^2}{T_\chi E_\chi(q)}\big)\frac{T_\chi-T_n}{T_\chi}e^{-\frac{E_\chi(q)-\mu_\chi}{T_\chi}}$. The $(T_\chi-T_n)$ on the numerator eventually becomes the temperature difference that is factorized out from the definition of kinetic energy transfer rate $\Gamma$~\cite{Visinelli:2015eka}. And the integration $\int\frac{d^3q}{(2\pi)^3}$ induces a factor of $\frac{n_\chi}{2S_\chi+1}$, which should be divided out for the rate of a single interested particle. Eventually
\begin{align}
\Gamma=&\sum_i\Big(\frac{\Gamma(\frac{i}{2}\hspace{-0.3em}+\hspace{-0.2em}3)(2m_nT_n)^{\frac{i+1}{2}}K_{\frac{i+5}{2}}({\textstyle\frac{m_n}{T_n}})}
{24\pi^{\frac{3}{2}}(2S_\chi\hspace{-0.3em}+\hspace{-0.2em}1)K_2({\textstyle\frac{m_n}{T_n}})}|\mathcal{M}|_i^2\Big)\frac{n_{n0}g_{\ast s}(T_n)T_n^3}{g_{\ast s0}T_\text{CMB}^3}\\
\times&\frac{1}{\frac{1}{3\pi^2}\frac{\Gamma(\frac{5}{2})}{\sqrt{\pi}}(2m_\chi T_\chi)m_\chi K_2({\textstyle\frac{m_\chi}{T_\chi}})e^{\frac{\mu_\chi}{T_\chi}}}\int\frac{d^3q}{(2\pi)^3}\frac{1}{E_\chi(q)^3}\frac{q^2}{3E_\chi(q)}\Big(3-\frac{q^2}{T_\chi E_\chi(q)}\Big)\frac{1}{T_\chi}e^{-\frac{E_\chi(q)-\mu_\chi}{T_\chi}},
\end{align}
which gives Eq.~\ref{eq:Gamma}.

\bibliography{WDMDoorFrame}

\end{document}